\documentclass[floats,floatfix,superscriptaddress,preprintnumbers,showpacs,eqsecnum,nofootinbib,notitlepage]{revtex4-1}

\usepackage[utf8]{inputenc}
\usepackage{latexsym,array,theorem,mathrsfs,bm,float}
\usepackage{psfrag}
\usepackage{amsfonts,amsmath,amssymb,latexsym,array,afterpage,
theorem,mathrsfs,bm,float,epsfig,color,graphicx,epstopdf,tabularx,here,multirow}

\begin{document}
\baselineskip=12pt
\newcommand{\nn}{\nonumber \\}
\newcommand{\D}{\nabla}

\preprint{WU-AP/1806/18, YITP-18-108, IPMU18-0158}

\title{Phenomenology in type-I minimally modified gravity}
\author{Katsuki \sc{Aoki}}
\email{katsuki-a12@gravity.phys.waseda.ac.jp}
\affiliation{Department of Physics, Waseda University, Shinjuku, Tokyo 169-8555, Japan}
\author{Antonio \sc{De Felice}}
\email{antonio.defelice@yukawa.kyoto-u.ac.jp}
\affiliation{Center for Gravitational Physics, Yukawa Institute for Theoretical Physics, Kyoto University, 606-8502, Kyoto, Japan}
\author{Chunshan \sc{Lin}}
\email{Chunshan.Lin@fuw.edu.pl}
\affiliation{Institute  of  Theoretical  Physics,  Faculty  of  Physics, University  of  Warsaw,  ul.  Pasteura  5,  Warsaw,  Poland}
\author{Shinji \sc{Mukohyama}}
\email{shinji.mukohyama@yukawa.kyoto-u.ac.jp}
\affiliation{Center for Gravitational Physics, Yukawa Institute for Theoretical Physics, Kyoto University, 606-8502, Kyoto, Japan}
\affiliation{Kavli Institute for the Physics and Mathematics of the Universe (WPI), The University of Tokyo, 277-8583, Chiba, Japan}
\affiliation{Institut Denis Poisson, UMR - CNRS 7013, Universit\'{e} de Tours, Parc de Grandmont, 37200 Tours, France}
\author{Michele \sc{Oliosi}}
\email{michele.oliosi@yukawa.kyoto-u.ac.jp}
\affiliation{Center for Gravitational Physics, Yukawa Institute for Theoretical Physics, Kyoto University, 606-8502, Kyoto, Japan}
\date{\today}

\begin{abstract}
  We study cosmology in a class of minimally modified gravity (MMG) with two local gravitational degrees of freedom. We classify modified gravity theories into type-I and type-II: theories of type-I have an Einstein frame and can be recast by change of variables as general relativity (GR) with a non-minimal matter coupling, while theories of type-II have no Einstein frame. Considering a canonical transformation of the lapse, the 3-dimensional induced metric and their conjugate momenta we generate type-I MMG. We then show that phenomenological deviations from GR, such as the speed of gravitational waves $c_T$ and the effective gravitational constant for scalar perturbations $G_{\rm eff}$, are characterized by two functions of an auxiliary variable. We study the phenomenology of several models all having $c_T=1$. We obtain a scenario with $c_T=1$ in which the effective equation-of-state parameter of dark energy is different from $-1$ even though the cosmic acceleration is caused by a bare cosmological constant, and we find that it is possible to reconstruct the theory on choosing a selected time-evolution for the effective dark energy component.
\end{abstract}



\maketitle

\section{Introduction}

There are (at least) three reasons for modifying gravity. First, it is the mysteries in the Universe such as dark energy, dark matter, inflation and big-bang singularity that motivate us to modify gravity. If we can replace some of those mysteries by relevant modifications of gravity then one might be able to find some clues to understand/solve them. Second, modification of gravity may help in constructing a theory of quantum gravity. We tend to think that general relativity (GR) should be modified at short distances for its reconciliation with quantum mechanics. Such examples include superstring theory, Ho\v{r}ava-Lifshitz gravity, ghost-free nonlocal gravity and so on. Third, modification of gravity may help us understand GR itself. Even if GR is the correct description of gravity, the only way to prove it (within some accuracy) is to constrain deviations from GR by observations and experiments. For this purpose we need theoretical models of possible modifications to compare with GR. Even from purely theoretical viewpoints, modification of gravity may lead to a better understanding of GR since one of the best ways to understand something is to break and reconstruct it. 

The number of local physical degrees of freedom in GR in $4$-dimensions is two, corresponding to the two polarizations of gravitational waves (GWs). Since we know that GWs exist in the realm of nature, the minimal number of local physical degrees of freedom in modified gravity should also be two. In order to minimally modify GR, one might thus wonder if it is possible to saturate this minimal number. 

In the search for theories of modified gravity that saturate the minimal number of local physical degrees of freedom ($=2$), one of the obstacles is the Lovelock theorem, which states that GR is unique if we assume the following four assumptions: (i) the number of spacetime dimensions is $4$; (ii) the theory is invariant under $4$-dimensional diffeomorphism; (iii) gravity is described by a metric only; and (iv) equations of motion of the form $E_{ab}=0$ include only up to second-order derivatives. In the present paper we consider theories that break (ii). This is in accord with the fact that any cosmological background breaks the $4$-dimensional diffeomorphism invariance. 

A class of minimally modified gravity (MMG) theories with $2$ local physical degrees of freedom was studied in \cite{Lin:2017oow}. Starting with $4$-dimensional metric theories invariant under $3$-dimensional diffeomorphisms and assuming as an ansatz that the action is linear in the lapse function, the necessary and sufficient condition under which a theory in this class has $2$ or less local physical degrees of freedom was found. Some simple examples with $2$ local physical degrees of freedom were also shown. However, it was not clear how to couple matter fields to gravity in a consistent way. Ref.~\cite{Aoki:2018zcv} then developed a general prescription to introduce matter fields to theories with $2$ local physical degrees of freedom. The purpose of the present paper is therefore to apply the general prescription to a class of theories that is wider than the one considered in \cite{Aoki:2018zcv} and then to study their phenomenology. 

The rest of the present paper is organized as follows. In section~\ref{sec:typeItypeII}, we classify modified gravity theories into type-I and type-II, and then introduce type-I MMG and type-II MMG. In section~\ref{sec:canonical}, we perform a canonical transformation to generate type-I MMG. We then study the cosmological dynamics and the scalar perturbations without imposing the constraint on the speed of GWs in section~\ref{sec:cosmology_general}. The case with $c_T=1$ is discussed in section~\ref{sec:cosmology_cT} in which we provide two concrete phenomenological scenarios of dark energy and a way to reconstruct the theory from a given cosmological dynamics. We make summary remarks in the last section~\ref{sec:summary}.

\section{Type-I and type-II minimally modified gravity}
\label{sec:typeItypeII}

In order to understand matter coupling in modified gravity theories in general, let us consider a scalar-tensor theory. In the Jordan (or matter) frame, the action including matter fields is of the form,
\begin{equation}
 I = \frac{1}{2}\int d^4x \sqrt{-g^{\rm J}}\left[ \Omega^2(\phi) R[g^{\rm J}] + \cdots \right]
  + I_{\rm matter}[g^{\rm J}_{\mu\nu}; {\rm matter}]\,, \label{eqn:scalartensor-Jordan}
\end{equation}
where the dots $\cdots$ include kinetic terms for $\phi$, and matter fields directly couple to the metric $g^{\rm J}_{\mu\nu}$. On the other hand, in the Einstein frame, $g^{\rm E}_{\mu\nu} = \Omega^2(\phi) g^{\rm J}_{\mu\nu}$\cite{Maeda:1988ab}, the action is of the form,
\begin{equation}
 I = \frac{1}{2}\int d^4x \sqrt{-g^{\rm E}}\left[ R[g^{\rm E}] + \cdots \right]
  + I_{\rm matter}[\Omega^{-2}(\phi)g^{\rm E}_{\mu\nu}; {\rm matter}]\,,
\end{equation}
where the gravity part is simply the Einstein-Hilbert action but matter fields couple to both the scalar field $\phi$ and the metric $g^{\rm E}_{\mu\nu}$ through the combination $\Omega^{-2}(\phi)g^{\rm E}_{\mu\nu}$. In the latter picture, although the gravity part of the action is exactly the same as that in GR, gravity is modified because of the non-trivial matter coupling. Thus the theory (\ref{eqn:scalartensor-Jordan}) can be recast as GR plus $\phi$ and matter fields. Let us call this type of modified gravity theories type-I. On the other hand, there are more general modified gravity theories in which there is no Einstein frame. Let us call them type-II.

{\it Type-I modified gravity} theories are those in which there exists an Einstein frame and which can be recast as GR plus extra degree(s) of freedom and matter fields by change of variables. In the Einstein frame, matter fields couple to gravity in non-trivial ways. On the other hand, {\it type-II modified gravity} theories are those in which there is no Einstein frame and which cannot be recast as GR plus extra degree(s) of freedom and matter fields by any change of variables.

{\it Type-II MMG} theories are therefore those metric theories of modified gravity with $2$ local physical degrees of freedom (= 2 polarizations of
transverse-traceless gravitational waves) in which there is no Einstein frame. By definition a type-II MMG theory cannot be recast as GR plus matter fields by any change of variables. A known example of type-II MMG is the minimal theory of massive gravity (MTMG) developed in \cite{DeFelice:2015hla,DeFelice:2015moy}~\footnote{The Lagrangian of MTMG includes some non-dynamical variable but one can integrate them out by using their equations of motion to rewrite the Lagrangian in terms of the metric only.}. The MTMG has $2$ local physical degrees of freedom corresponding to massive GWs, admits a self-accelerating cosmological solution, and is free from instabilities such as Boulware-Deser ghost~\cite{Boulware:1973my}, Higuchi ghost~\cite{Higuchi:1986py} and nonlinear ghost~\cite{DeFelice:2012mx}. Also, the recently developed positivity bound~\cite{Cheung:2016yqr,Bonifacio:2016wcb,Bellazzini:2017fep,deRham:2017xox,deRham:2018qqo} that significantly restricts the viability of Lorentz-invariant massive gravity theories does not apply to the MTMG. Thanks to the absence of extra degrees of freedom, MTMG also enjoys interesting phenomenology~\cite{DeFelice:2016ufg,Bolis:2018vzs,DeFelice:2018vza,Fujita:2018ehq}.

{\it Type-I MMG} theories are, on the other hand, those metric theories of modified gravity with $2$ local physical degrees of freedom (= 2 polarizations of
transverse-traceless gravitational waves) in which there exists an Einstein frame. A type-I MMG in the absence of matter fields is equivalent to GR, but gravity is modified because of non-trivial matter coupling. A systematic study of type-I MMG was initiated in \cite{Aoki:2018zcv}. By definition a type-I MMG can be recast as GR plus matter fields, which couple non-trivially, by change of variables. Since the most general change of variables is a canonical transformation, Ref.~\cite{Aoki:2018zcv} started with GR and applied a canonical transformation to the ADM variables and their conjugate momenta. If we simply add matter fields to the system after the canonical transformation then the setup turns out to be inconsistent. This is because the addition of matter fields breaks the time diffeomorphism invariance and thus downgrades one of the first-class constraints to second-class, leaving an extra degree of freedom in the phase space. Ref.~\cite{Aoki:2018zcv} then found a consistent way of matter coupling: after a canonical transformation, one needs to impose a gauge-fixing condition as an additional constraint before adding matter fields. In this way, the first-class constraint associated with the time diffeomorphism invariance is split into a pair of second-class constraints and the pair remains second-class after the addition of matter fields. This prescription leads to explicit examples of metric theories of gravity with $2$ local physical degrees of freedom that can be recast as GR plus matter fields, which couple non-trivially. Therefore this is an explicit and general construction of type-I MMG theories.

As pointed out in \cite{Aoki:2018zcv}, the same prescription can be applied also to the examples of MMG in \cite{Lin:2017oow}: one can simply add matter fields after gauge-fixing. Ref.~\cite{Carballo-Rubio:2018czn} argued that some of the examples of MMG considered in \cite{Lin:2017oow}, such as the square-root gravity, should be equivalent to GR up to change of variables in the absence of matter field. In other words, those examples should be of type-I, provided that matter fields can be consistently coupled to gravity. While Ref.~\cite{Carballo-Rubio:2018czn} unfortunately failed to find a consistent matter coupling, one can easily apply the prescription of Ref.~\cite{Aoki:2018zcv} to those examples in \cite{Lin:2017oow} to establish a class of type-I MMG. 

While the prescription of \cite{Aoki:2018zcv} is completely general and can be applied to any type-I MMG theories, a class of canonical transformations studied there was not general. In the rest of the present paper we thus consider a more general class of canonical transformations and then study the phenomenology of the correspondingly more general class of type-I MMG theories.

\section{Canonical transformation including lapse function}
\label{sec:canonical}
The gravitational Hamiltonian is given by
\begin{align}
H_{\rm tot}=\int d^3x (\mathcal{N} \mathcal{H}_0[\Gamma,\Pi] + \mathcal{N}^i \mathcal{H}_i[\Gamma,\Pi] + \lambda \Pi_N +\lambda^i \Pi_i ) \,,
\end{align}
with 
\begin{align}
\mathcal{H}_0 &:=\frac{2}{M^2 \sqrt{\Gamma}}\left( \Gamma_{i k} \Gamma_{j l} -\frac{1}{2}\Gamma_{ij} \Gamma_{kl} \right) \Pi^{ij}\Pi^{kl}
-\frac{M^2 \sqrt{\Gamma}}{2} R(\Gamma)\,, \\
\mathcal{H}_i &:= -2\sqrt{\Gamma} \Gamma_{ij} D_k \left( \frac{\Pi^{jk}}{\sqrt{\Gamma}} \right) \,, 
\end{align}
where $M$ is a mass scale, and where $(\mathcal{N},\Pi_\mathcal{N})$, $ (\mathcal{N}^i,\Pi_i)$, and $(\Gamma_{ij},\Pi^{ij})$ are the ADM variables and their canonical conjugates of the Einstein frame metric:
\begin{align}
ds_{\rm E}^2=g^{\rm E}_{\mu\nu}dx^{\mu}dx^{\nu}=-\mathcal{N}^2dt^2+\Gamma_{ij}(dx^i+\mathcal{N}^i dt)(dx^j+\mathcal{N}^j dt)\,.
\end{align}
On the other hand, we assume that matter fields minimally couple with the Jordan frame metric,
\begin{align}
ds_{\rm J}^2=g^{\rm J}_{\mu\nu}dx^{\mu}dx^{\nu}=-N^2dt^2+\gamma_{ij}(dx^i+N^i dt)(dx^j+N^j dt)\,.
\end{align}
and that the two metrics are related by a canonical transformation, $(\mathcal{N},\Pi_\mathcal{N},\mathcal{N}^i,\Pi_i, \Gamma_{ij},\Pi^{ij}) \rightarrow (N,\pi_N,N^i,\pi_i, \gamma_{ij},\pi^{ij})$. Although one may discuss a quite general canonical transformation, just for simplicity, we consider the case in which only the lapse, the spatial metric and their conjugate momenta change via the canonical transformation and in which the generating functional is given by
\begin{align}
F=-\int d^3x ( M^2 \sqrt{\gamma} f(\tilde{\Pi},\tilde{\mathcal{H}}) + N^i \Pi_i )
\end{align}
where
\begin{align}
\tilde{\Pi}=\frac{1}{M^2 \sqrt{\gamma}}\Pi^{ij}\gamma_{ij} 
\,, ~
\tilde{\mathcal{H}}=\frac{1}{M^2 \sqrt{\gamma}}\Pi_\mathcal{N} N\,,
\end{align}
and $f$ is an arbitrary function of $\tilde{\Pi}$ and $\tilde{\mathcal{H}}$.
Then, we obtain
\begin{align}
\Gamma_{ij}&=-\frac{\delta F}{\delta \Pi^{ij}}=f_{\tilde{\Pi}} \gamma_{ij} \,,
\\
\mathcal{N}&=-\frac{\delta F}{\delta \Pi_\mathcal{N}}=f_{\tilde{\mathcal{H}}}N \,, \\
\pi^{ij}&=-\frac{\delta F}{\delta \gamma_{ij}}
=f_{\tilde{\Pi}}\Pi^{ij}+\frac{M^2 }{2}\sqrt{\gamma} \gamma^{ij }\left( f - f_{\tilde{\Pi}} \tilde{\Pi}- f_{\tilde{\mathcal{H}}} \tilde{\mathcal{H}} \right)
\,,\\
\pi_N&=-\frac{\delta F}{\delta N}=f_{\tilde{\mathcal{H}}} \Pi_\mathcal{N}\,,
\end{align}
and
\begin{equation}
 \mathcal{N}^i = N^i\,, \quad \pi_i = \Pi_i\,,
\end{equation}
with the notations $f_{\tilde{\Pi}}=\partial f /\partial \tilde{\Pi}$ and $f_{\tilde{\mathcal{H}}}=\partial f /\partial \tilde{\mathcal{H}}$.
Introducing auxiliary variables $\phi$ and $\psi$, the old variables are given by
\begin{align}
\Gamma_{ij}&=f_{\phi}\gamma_{ij} \,,
\\
\Pi^{ij}&=\frac{1}{f_{\phi}} \left[ \pi^{ij}-\frac{M^2}{2}\sqrt{\gamma}(f-f_{\phi}\phi - f_{\psi} \psi) \gamma^{ij} \right]
\,, \\
\mathcal{N}&=f_{\psi}N
\,, \\
\Pi_\mathcal{N}&=\frac{1}{f_{\psi}}\pi_N\,,
\end{align}
and
\begin{equation}
 \mathcal{N}^i = N^i\,, \quad \Pi_i = \pi_i\,,
\end{equation}
where $\phi$ and $\psi$ satisfy the constraints
\begin{align}
\mathcal{C}:&=\pi^{ij}\gamma_{ij}-\frac{M^2}{2}\sqrt{\gamma}(3f-f_{\phi}\phi)+\frac{3M^2}{2}\sqrt{\gamma} f_{\psi} \psi \approx 0
\,, \\
\mathcal{D}:&=\pi_N N-M^2\sqrt{\gamma} f_{\psi} \psi \approx 0 
\,.
\end{align}

We then rewrite the Hamiltonian in terms of the new variables $(\gamma_{ij},\pi^{ij},N,\pi_N)$.
Since $\phi$ and $\psi$ are non-dynamical variables, after introducing the Lagrange multipliers to implement the constraints on the canonical pairs $(\phi,\pi_{\phi})$ and $(\psi,\pi_{\psi})$, the Hamiltonian in terms of the new variables is given by
\begin{align}
H_{\rm tot}=\int d^3x (N f_{\psi} \mathcal{H}_0+N^i \mathcal{H}_i+
 \lambda \pi_N +\lambda^i \pi_i +\lambda_C \mathcal{C} +\lambda_D \mathcal{D} +\lambda_{\phi} \pi_{\phi}+\lambda_{\psi} \pi_{\psi} ) \,,
\end{align}
where
\begin{align}
\mathcal{H}_0&=\frac{2}{M^2 f_{\phi}^{3/2} \sqrt{\gamma}}\left( \gamma_{ik}\gamma_{jl}-\frac{1}{2}\gamma_{ij}\gamma_{kl} \right)
\left[ \pi^{ij}-\frac{M^2}{2}\sqrt{\gamma}(f-f_{\phi}\phi-f_{\psi}\psi) \gamma^{ij} \right]
\left[ \pi^{kl}-\frac{M^2}{2}\sqrt{\gamma}(f-f_{\phi}\phi-f_{\psi}\psi) \gamma^{kl} \right]
\nn
&-\frac{M^2 f_{\phi}^{1/2} \sqrt{\gamma}}{2}\left[ R[\gamma]-2 \D^2 \ln f_{\phi} -\frac{1}{2} \D_i \ln f_{\phi} \D^i \ln f_{\phi} \right] , \\
\mathcal{H}_i &=-2\sqrt{\gamma} \gamma_{ij}\D_k \left(\frac{\pi^{jk}}{\sqrt{\gamma}} \right)+\mathcal{C} \partial_i \ln f_{\phi} + 
\mathcal{D} \partial_i \ln f_{\psi} - \pi_N N \partial_i \ln f_{\psi}
\,,
\end{align}
and $\D_i$ is the covariant derivative with respect to $\gamma_{ij}$. 
Note that the primary constraint $\pi_N \approx 0$ leads to $\psi \approx 0$ due to $\mathcal{D} \approx 0$. Since we obtain $\psi \approx 0, \pi_{\psi} \approx 0$, we can eliminate the pair $(\psi,\pi_{\psi} )$ from the dynamical variables. Hence, it is useful to expand $f$ in terms of $\psi$,
\begin{align}
f(\phi,\psi)=f_0 (\phi)+f_1 (\phi) \psi +\mathcal{O}(\psi^2) \,.
\end{align}
We find that higher order terms of $\psi$ do not appear in the Hamiltonian when imposing $\psi \approx 0$ and thus it is sufficient to expand $f$ up to linear order in $\psi$.
Redefining the Lagrange multipliers, we then obtain
\begin{align}
H_{\rm tot}=\int d^3x (N \mathcal{H}'_0[\gamma,\pi,\phi ]+N^i \mathcal{H}'_i[\gamma, \pi]+
 \lambda \pi_N +\lambda^i \pi_i +\lambda_C \mathcal{C}' +\lambda_{\phi} \pi_{\phi} )\,, \label{Htot}
\end{align}
where
\begin{align}
\mathcal{H}'_0 &:=
\frac{2f_1 }{f_0'{}^{3/2}}\left[ \frac{1}{M^2 \sqrt{\gamma}}\left(\gamma_{ik}\gamma_{jl}-\frac{1}{2}\gamma_{ij}\gamma_{kl}\right)\pi^{ij}\pi^{kl}
+\frac{M^2\sqrt{\gamma}}{8}(f_0-f_0'\phi)(3f_0+f_0'\phi) \right]
\nonumber \\
&-\frac{f_1 M^2 \sqrt{\gamma}}{2}\sqrt{f_0'} \left[ R(\gamma)-2\frac{f_0''}{f_0'}\nabla^2\phi -\left( \frac{\nabla_i \phi }{f_0'} \right)^2
\left( 2f_0'f_0'''-\frac{3}{2}f_0''{}^2 \right)
\right] , \label{HC}
\\
\mathcal{H}'_i &:=-2\sqrt{\gamma} \gamma_{ij}\D_k\! \left(\frac{\pi^{jk}}{\sqrt{\gamma}} \right)
, \\
\mathcal{C}'&:=\pi^{ij}\gamma_{ij}-\frac{M^2}{2}\sqrt{\gamma}(3f_0-f_0' \phi)
\,,
\end{align}
with $f'_0=df_0 /d\phi$, $f_0''=d^2f_0 /d\phi^2$ and $f'''_0=d^3f_0 / d\phi^3$. We require $f_0'>0$ and $f_1>0$ so that positive definite $\Gamma_{ij}$ and $\mathcal{N}$ result in positive definite $\gamma_{ij}$ and $N$. This will also prevent the apparition of any ghosts in the tensor sector.

Since the canonical transformation does not change the structure of the Hamiltonian, \eqref{Htot} has eight first class constraints as is the case in GR. In addition, \eqref{Htot} has two second class constraints associated with the canonical pair $(\phi, \pi_{\phi})$. The number of degrees of freedom of \eqref{Htot} is thus four in the phase space. However, that is not the case when a matter field is introduced as shown in \cite{Aoki:2018zcv}. In such a case, a (previously) first class constraint associated with the temporal diffeomorphism invariance turns out to be second class due to the matter coupling. As a consequence, in general, one extra mode (besides those of the matter field) appears in the phase space. A consistent way to introduce the matter field is, before inclusion of the matter fields, to split the first class constraint into a pair of second class constraints by introducing a ``gauge fixing condition.'' Since these constraints remain second class after introducing the matter field, the number of gravitational degrees of freedom remains four in the phase space, i.e., two in the real space.

Therefore, the final expression of the canonically-transformed and gauge-fixed Hamiltonian is 
\begin{align}
H_{\rm tot}=\int d^3x (N \mathcal{H}'_0+N^i \mathcal{H}'_i+
 \lambda \pi_N +\lambda^i \pi_i +\lambda_C \mathcal{C}' +\lambda_{\phi} \pi_{\phi} 
+ \sqrt{\gamma}\,  \mathcal{H}_{\rm gf}) \,, 
\end{align}
with the gauge fixing terms $\mathcal{H}_{\rm gf}$. The gauge fixing term is actually part of the definition of the theory. Each inequivalent gauge fixing will lead, in general, to a different theory. When the gauge fixing term does not contain $\pi^{ij}$, taking the Legendre transformation and integrating out $\lambda_C$, we obtain the Lagrangian
\begin{align}
N \sqrt{\gamma} \mathcal{L}&= M^2 f_1 N \sqrt{\gamma} \Biggl[ \frac{f_0'{}^{3/2}}{2 } \left(  \frac{1}{f_1^2} K^{ij}K_{ij}-\frac{1}{3f_1^2 }K^2 +\frac{1 }{f_0'} R(\gamma) \right)
+\frac{K}{f_1} \left( f_0-\frac{1}{3}f_0'\phi \right)
\nonumber \\
&\qquad \qquad \qquad \quad
-\frac{ f_0''}{f_0'{}^{1/2}}\nabla^2\phi -\left( \frac{\nabla_i \phi }{f_0'{}^{3/4}} \right)^2 \left(f_0'f_0'''-\frac{3}{4}f_0''{}^2 \right)+\frac{1}{3}f_0'{}^{1/2}\phi^2 
\Biggl] - \sqrt{\gamma}\, \mathcal{H}_{\rm gf} \,,
\label{action_with_Phi}
\end{align}
where $K_{ij}$ is the extrinsic curvature, $K_{ij}:= (\dot{\gamma}_{ij}-2\D_{(i} N_{j)} )/2N$.

In the present paper, we consider
\begin{align}
\mathcal{H}_{\rm gf}=\lambda_{\rm gf}^i \partial_i \phi \,,\label{eq:gauge-fix}
\end{align}
with the Lagrange multiplier $\lambda_{\rm gf}^i$ transforming the first class constraint associated with the temporal diffeomorphism invariance into a pair of second class constraints\footnote{One may also impose the same gauge condition by $\mathcal{H}_{\rm gf}=\lambda_{\rm gf} \D^2 \phi$ with an appropriate boundary condition on $\phi$.}. It is worthwhile mentioning here that $\phi$ is not a scalar with respect to the $4$-dimensional diffeomorphism although it is a scalar with respect to the $3$-dimensional diffeomorphism. At the level of the Hamiltonian, on using the gauge-fixing term (\ref{eq:gauge-fix}), we can make a field redefinition of the Lagrange multiplier $\lambda_{\rm gf}^i$ in order to remove all the gradient or Laplacian terms of the field $\phi$ from the Hamiltonian written in Eq.\ (\ref{Htot}). Then, we can show that the gauge-fixed Lagrangian is given by
\begin{align}
N\sqrt{\gamma} \mathcal{L}= M^2 f_1 N \sqrt{\gamma} \Biggl[ \frac{f_0'{}^{3/2}}{2 } \left(  \frac{1}{f_1^2} K^{ij}K_{ij}-\frac{1}{3f_1^2 }K^2 +\frac{1 }{f_0'} R(\gamma) \right)
+\frac{K}{f_1} \left( f_0-\frac{1}{3}f_0'\phi \right)
+\frac{1}{3}f_0'{}^{1/2}\phi^2 
 \Biggl] - \sqrt{\gamma}\lambda_{\rm gf}^i \partial_i \phi\,, \label{eqn:gaugefixedaction}
\end{align}
and it respects both the time-reparametrization symmetry and the spatial diffeomorphism invariance,
\begin{align}
t\rightarrow t'(t)\,, \quad x^i \rightarrow x'{}^i(t,x^i)\,.
\end{align}
Generically, given functions $f_0$ and $f_1$, $\phi$ may be integrated out. In the general case, however, it is more convenient to work with an auxiliary variable, which prevents cumbersome calculations. 

From the Lagrangian it is obvious that the squared speed of GWs is 
\begin{align}
c_T^2=f_1^2/f_0' \,.
\end{align}
On the other hand, once we define this frame as the Jordan frame, the speed of light is given by the usual value\textemdash here $c_\text{EM} = 1$. Hence, unity speed for GWs is realized when $f_0'=f_1^2$. It is in principle possible to see the same difference arising in the Einstein frame  due to the modification of the matter coupling.

In \cite{Aoki:2018zcv}, a less general canonical transformation was considered, and in that case a theory, whenever $c_T^2=1$, would necessarily coincide with GR. In the present work, due to the more general canonical transformation, the theories do not correspond to GR generally even on setting unity speed for GWs, since there is still some freedom to set the functions $f_0$ and $f_1$. This is made explicit in the following sections. In fact, in the following, we will consider only theories for which, at least at late times, we have that $c_T=1$. In this case the Jordan frame metric and the Einstein frame metric are conformally related by
\begin{align}
g^{\rm J}_{\mu\nu}=f_0'(\phi) g^{\rm E}_{\mu\nu}\,. \label{conf}
\end{align}
We emphasize that this is not the usual conformal transformation $g_{\mu\nu} \rightarrow \Omega^2(\phi)g_{\mu\nu}$ with an independent scalar variable $\phi$ because the variable $\phi$ in \eqref{conf} is not independent from the metric. The auxiliary variable $\phi$ is related to the metric (and its conjugate momentum) via the constraint $\mathcal{C}' \approx 0$. Moreover, $\phi$ is not a $4$d scalar although it is a $3$d scalar. 

The limit $f_1\rightarrow 1$, $f_0' \rightarrow 1$ gives a GR limit in the uniform Hubble slicing $K=K(t)$ because the equation of motion of $\phi$ yields $\phi=-K$. We notice that $f_1\rightarrow M_*/M$, $f_0' \rightarrow M_*^2/M^2$ with a constant $M_*$ also gives another GR limit with a new mass scale $M_*$.

\section{Cosmology for general $\lowercase{f}_0(\phi)$ and $\lowercase{f}_1(\phi)$}
\label{sec:cosmology_general}
In this section, we shall consider the dynamics of the FLRW universe with a perfect fluid and the scalar perturbations around it for a general choice of $f_0(\phi)$ and $f_1(\phi)$, in particular without imposing the condition $c_T^2=1$. This case is meant to describe the universe at early times, where the condition $c_T^2=1$ does not have to be satisfied. The specific cases with $c_T^2=1$ will be discussed in the next section. To represent a perfect fluid, we introduce the $k$-essence field
\begin{align}
S_{\rm m}=\int d^4 x \sqrt{-g^{\rm J}} P(X) \,,
\end{align}
where
\begin{align}
X=-\frac{1}{2}g^{{\rm J} \mu\nu} \partial_{\mu} \sigma \partial_{\nu} \sigma \,.
\end{align}
For such a fluid, we can write its energy density and pressure as
\begin{align}
\rho=-P+2XP_X\,, \quad p= P\,,
\end{align}
where $P_X=dP/dX$ and $P_{XX}=d^2P/dX^2$. As already mentioned, we assume the gauge condition imposed by (\ref{eq:gauge-fix}) throughout. 

Once matter is added $f_0'^{3/2}/f_1$ becomes the no-ghost condition for the tensor modes. This is however already satisfied, since both free functions were already chosen as positive.

For the background metric, on using the time-reparametrization invariance to simplify the lapse function, we can write:
\begin{align}
N=1 \,, \quad \gamma_{ij}=a^2(t) \delta_{ij}\,,
\end{align}
and the background $k$-essence field
\begin{align}
\sigma=\bar{\sigma}(t)\,.
\end{align}
Then, we have following three independent equations
\begin{align}
\frac{M^2}{3} \phi^2 f_1 \sqrt{f_0'} &= \bar{\rho}
\,, \label{eq_phi}
\\
6H\sqrt{f_0'} f_{0m}&=-\phi  d_f \,, \label{eq_H}
\end{align}
and
\begin{align}
\dot{\bar{\rho}}+3H (\bar{\rho}+\bar{p})=0 \,, \label{eqn:conservation}
\end{align}
where $H:=\dot{a}/a$ is the Hubble expansion rate and
\begin{align}
f_{0m}&:=2f_0'-f_0'' \phi \,, \\
d_f&:= 4 f_0' f_1 +f_0'' f_1 \phi +2 f_0' f_1' \phi \,.
\end{align}
Throughout the present paper, we use the bar to represent the respective background quantity.
From the above two equations (\ref{eq_phi}) and (\ref{eq_H}), we obtain the Friedmann equation
\begin{align}
3M^2H^2= \bar{\rho} + \frac{M^2 \phi^2 }{12}
\left[ -4 f_1 \sqrt{f_0'}+ \frac{d_f^2}{f_0' f_{0m}^2} \right] \,. \label{eqn:Friedmann-eq}
\end{align}
Notice here that, on the background, we have that the gravitation constant can be identified with $1/(8\pi M^2)$. As we will see later on, this value does not coincide, in general, with the expression of the effective gravitational constant, $G_{\rm eff}$, which drives the linear dynamics of the dark matter fluid fluctuations.

We then consider the scalar perturbations
\begin{align}
N&=1+\alpha(t,\mathbf{x})  \,, \nn 
N^i&=\delta^{ij} \partial_j \beta (t,\mathbf{x}) \,, \nn
\gamma_{ij}&=a^2\,[1+2\zeta(t,\mathbf{x}) ] \delta_{ij}\,,
\end{align}
and
\begin{align}
\sigma=\bar{\sigma}+\delta \sigma(t,\mathbf{x})
\,,
\end{align}
where we have fixed the spatial gauge so that the spatial metric becomes diagonal. We define the gauge invariant variables
\begin{align}
\Psi &:= \alpha +\frac{\partial}{\partial t}(a^2 \beta)
\,, \\
\Phi&:=-\zeta-a^2 H \beta
\,, \\
\delta &:= \frac{\delta \rho}{\bar{\rho}} +3(1+\bar{p}/\bar{\rho}) \zeta\,,
\end{align}
where
\begin{eqnarray}
 \delta\rho & = & \frac{\bar{\rho}+\bar{p}}{c_{\sigma}^2}\left(\frac{\partial_t\delta\sigma}{\partial_t\bar{\sigma}}-\alpha\right)\,, \nonumber\\
  c_{\sigma}^2 & = & \frac{P_X}{P_X+2XP_{XX}}\,, 
\end{eqnarray}
and use $(\Phi,\Psi,\delta,\beta)$ as independent variables of perturbations instead of $(\alpha,\beta,\zeta,\delta\sigma)$. Note that the perturbations of the $k$-essence, in the dust limit $\bar{p} \rightarrow 0$, $ c_{\sigma}^2 \rightarrow 0$, are singular when we use $\delta \sigma$ as the independent variable, while a regular dust limit can be taken safely when $\delta$ is used for the independent variables of the perturbations (see Appendix of \cite{DeFelice:2015moy}). After integrating out the non-dynamical variables $(\Phi,\Psi,\beta)$, we obtain the quadratic order action for the density perturbations
\begin{align}
  N\sqrt{\gamma} \mathcal{L}_{\delta}=\mathcal{A}\,\dot\delta^2-\mathcal{B}\,\delta^2\,,
\end{align}
in the momentum space. In the high-$k$ limit, we can write down the no-ghost condition
\begin{eqnarray}
  Q=\lim_{k\to\infty}\frac{\mathcal{A}}{a^3}=\frac12\,\frac{a^2}{k^2}\,\frac{\bar\rho^2}{\bar\rho+\bar p}\,,
\end{eqnarray}
Therefore, there is no ghost instability as long as $\bar{\rho}+\bar p>0$. This result is equivalent to the standard case in GR. Along the same lines we find that matter waves, in the high-$k$ limit, propagate with the squared-speed 
\begin{eqnarray}
  c_s^2=\lim_{k\to\infty}\frac{a^2}{k^2}\,\frac{\mathcal{B}}{\mathcal{A}}=c_{\sigma}^2\,. 
\end{eqnarray}
Therefore, in terms of speed of propagation, for both radiation and dust we have the same results as in GR. At even earlier times, say during BBN, baryons will be ultra-relativistic and in general coupled with photons, so that the assumption of single gravitating fluid will break down. It is of course easy and safe to couple multiple matter fields to the system, as far as they are added after the gauge-fixing. We nonetheless for simplicity focus in the following section on constraints we can firmly set on the late time behavior of the theory.

\section{Cosmology with $\lowercase{c}_T^2=1$}
\label{sec:cosmology_cT}
The recent observations of GWs have revealed that the speed of GWs is the same as the speed of light with a high degree of accuracy\footnote{By considering Hubble friction effects on the propagation, one may find additional less stringent constraints (see e.g. \cite{Belgacem2018}).} $(\lesssim 10^{-15})$ in the present universe~\cite{TheLIGOScientific:2017qsa,Monitor:2017mdv}. We thus consider the specific case $c_T^2=1$, i.e., $f_1^2=f_0'$, in this section and discuss some applications to the present accelerating expansion of the universe. Notice that this choice for the function $f_1$ will lead to the value of $c_T=1$ being valid at all times. From the coefficients of $K^{ij}K_{ij}$ and $R(\gamma)$ in (\ref{eqn:gaugefixedaction}) we see that the same value for $c_T$ will also hold on different backgrounds.

After setting $c_T=1$, the set of independent equations of motion for the flat FLRW background (\ref{eq_phi})-(\ref{eqn:conservation}) is
\begin{eqnarray}
M^2 f'_0 \phi^2  - 3\bar{\rho} & = & 0\,,\label{eqn:eq_f'0} \\
f''_0 \phi^2 + (2f'_0-3Hf''_0)\phi + 6 Hf'_0 & = & 0\,,\label{eqn:eq_phi} \\
 \dot{\bar{\rho}} + 3H(\bar{\rho}+\bar{P}) & = & 0\,.
\end{eqnarray}
By using the first and third equations, the second equation can be rewritten as
\begin{equation}
 \dot{\phi} + \frac{\bar{\rho}+\bar{P}}{4\bar{\rho}}(3H\phi-\phi^2) = 0\,.\label{eq_phi_cT1}
\end{equation}

For the purpose of realizing cosmic acceleration, we subtract a possibly present constant term from the total energy density and the total pressure,
\begin{align}
\rho&=M^2 \Lambda+\rho_{\rm m} \,, \\
p&=-M^2 \Lambda + p_{\rm m}\,,
\end{align}
where $\Lambda$ is the bare cosmological constant, and $\rho_{\rm m}$ and $p_{\rm m}$ are respectively the energy density and the pressure for a matter field. The Friedmann equation (\ref{eqn:Friedmann-eq}) is then 
\begin{align}
3M^2 H^2 = \bar{\rho}_{\rm m} +\bar{\rho}_{\rm DE}\,,
\end{align}
where
\begin{align}
\bar{\rho}_{\rm DE}:= M^2  \Lambda+ \frac{M^2 \phi^2 }{3}
\left[  \frac{(2f_0'+\phi f_0'' )^2 }{(2f_0' -\phi f_0'')^2} - f_0' \right] \,, \label{eqn:def-rhoDE}
\end{align}
is the effective energy density of the dark energy. The effective pressure of the dark energy is defined by
\begin{align}
\bar{p}_{\rm DE}:=-\bar{p}_{\rm m} -2 M^2 \dot{H}-\bar{\rho}_{\rm DE}-\bar{\rho}_{\rm m} \,.
\end{align}
Since in the following we consider only the evolution at late times, we will neglect the radiation component and we will suppose that beside a possible cosmological constant we have a cold dust fluid, with equations of state $p_{\rm m}=0$ and $T_{\rm m}=0$. In this case, the equation of state parameter of dark energy is then
\begin{align}
w_{\rm DE}&:=\frac{\bar{p}_{\rm DE}}{\bar{\rho}_{\rm DE}}
\nn
&\, = -
\frac{6\Lambda (4 f_0'{}^2 -5 \phi^2 f_0''{}^2+4 \phi f_0' f_0 '' +4 \phi^2 f_0' f_0''') + \phi^3 (8\phi f_0' f_0''{}^2 - \phi^2 f_0''{}^3 -4 f_0'{}^2 f_0 '' -8 \phi f_0'{}^2 f_0''')}{ (2f_0'-\phi f_0'')[ 3\Lambda (2f_0'-\phi f_0'')^2 - \phi^2 \{ 4(f_0'-1) f_0'{}^2 -4 \phi f_0' (f_0'+1) f_0'' +\phi^2 (f_0'-1) f_0''{}^2 \} ]}
\,.
\end{align}
We need to assume $w_{\rm DE}<-1/3$, at least at low redshifts, in order to have an accelerating universe.

On borrowing the results of the previous section, we find that the reduced action for the field $\delta$ of the dust fluid can be written as
\begin{align}
N\sqrt{\gamma} \mathcal{L}_{\delta}=\frac{1}{2}\,a^3 \bar{\rho}\,\bar{\rho}_{\rm m} \left[ \left( \frac{1}{2}\,\bar\rho_{\rm m}\phi^2 + \frac{k^2}{a^2}\,\bar\rho \right)^{-1} \dot{\delta}^2
+\frac{1}{6}\, \frac{k^2}{a^2}\,\bar\rho_{\rm m} \phi^2\left( \frac{1}{2}\bar\rho_{\rm m} \phi^2+\frac{k^2}{a^2}\,\bar\rho \right)^{-2} \delta^2
\right],
\end{align}
in the momentum space, where $\bar\rho := M^2 \Lambda+\bar\rho_{\rm m}$.

In the sub-horizon limit, the definition of $G_{\rm eff}$ is gauge independent and we can set constraints on it by using existing data (e.g.\ RSD data). In our case, in the sub-horizon limit, the equation of motion of the matter density perturbation leads to
\begin{align}
\ddot{\delta}+2 H \dot{\delta}-4\pi G_{\rm eff} \bar{\rho}_{\rm m}\, \delta =0 \,,
\end{align}
whereas the Poisson equation can be written as
\begin{equation}
-\frac{k^2}{a^2} \Psi= 4\pi G_{\rm eff} \bar{\rho}_{\rm m}\, \delta\,,
\end{equation}
where
\begin{equation}
8\pi M^2 G_{\rm eff} = \frac{1}{f_0'}\,. \label{eqn:Geff}
\end{equation}
In this same limit it is also possible to show that the slip parameter is given by
$\eta:=\Psi/\Phi=1$. These results immediately lead to the conclusion that the $G_{\rm eff}$ will tend to the GR limit in the case $f_0'\to1$. Later on, when we discuss explicit models, we will consider theories for which this GR-limit holds at early times. 

Furthermore, since the gravity at short scales is dominated by $G_{\rm eff}$, we will set that $G_{\rm eff}(z=0)=G_N$, where $z$ is the redshift and $G_N$ is the Newton gravitational constant. Therefore, this theory will have in general a time-dependent $G_{\rm eff}$, which will differ in the past, in general, from its today's value, namely $G_N$.

As we will show in the following subsection, the accelerating expansion of the universe can be realized even in the case $\Lambda=0$. However, we will also discuss a case with a non-zero cosmological constant. The latter case cannot explain the origin of the acceleration but it still allows the possibility of non-trivial time evolutions for $G_{\rm eff}$ as well as $w_{\rm DE}$, which, in turn, may give phenomenologically interesting scenarios. We thus discuss the cases with $\Lambda =0$ and with $\Lambda \neq 0$ in order.

\subsection{Models without $\Lambda$}

So far, we have considered the theory with a general function $f_0(\phi)$. The choice of such a function may lead in general to a non-trivial phenomenology. For example, in order to have a de Sitter solution in the case $\Lambda=0$, we first set a functional form of $f'_0$ by solving \eqref{eq_H} with $f_0'=f_1^2$ for $H=H_*=$ constant. The solution is
\begin{align}
f'_{0,{\rm dS}}=\frac{\phi_c^2 \phi^2}{(3H_*-\phi)^4}
\,, \label{fdS}
\end{align}
with a constant $\phi_c$. Note that \eqref{fdS} yields $H=H_*$ independently from values of $\phi$ and $\bar{\rho}$, and thus $f'_0=f'_{0,{\rm dS}}$ cannot give a viable cosmological solution. Besides, \eqref{eq_phi} implies $\phi \rightarrow 0$ for $\bar{\rho} \rightarrow 0$. Therefore, we assume $f'_0 \rightarrow f'_{0,{\rm dS}}$ as $\phi \rightarrow 0$ to have the late time accelerating expansion. Furthermore, we also assume $f'_0 \rightarrow 1$ as $|\phi| \rightarrow \infty$ in order to recover GR in the early stage of the universe. A simple function to satisfy such conditions is
\begin{align}
f_0'=\frac{\phi_c^2 \phi^2 +\phi^4}{(3H_* -\phi )^4} \,. \label{fdS2}
\end{align}
GR is recovered when $|\phi| \gg |\phi_c|$, while the modification from GR appears for $|\phi| \ll |\phi_c|$ and then the de Sitter expansion is obtained.

Substituting \eqref{fdS2} into \eqref{eq_phi} with $f_0'=f_1^2$ we obtain
\begin{align}
H=\frac{6H_* \phi_c^2 +9 H_* \phi^2 -\phi^3}{6 \phi_c^2 +9 H_* \phi +3 \phi^2} \,.
\end{align}
Since we have $\phi \rightarrow 0$ as $\bar{\rho}_{\rm m} \rightarrow 0$ and $\phi \rightarrow - \infty$ as $\bar{\rho}_{\rm m} \rightarrow \infty$, we need to assume $6 \phi_c^2 +9 H_* \phi +3 \phi^2 \neq 0$ for $-\infty < \phi <0$ in order not to have a singularity $H\rightarrow \infty$ at a finite value of $\bar{\rho}_{\rm m}$. The roots of $6 \phi_c^2 +9 H_* \phi +3 \phi^2 = 0$ are
\begin{align}
\phi = -\frac{1}{2} \left( 3 H_* \pm \sqrt{9H_*^2 -8 \phi_c^2} \right) \,.
\end{align}
Hence, the condition $6 \phi_c^2 +9 H_* \phi +3 \phi^2 \neq 0$ for $-\infty < \phi <0$ yields
\begin{align}
8\phi_c^2 > 9 H_*^2 \,.
\end{align}


We find that today typically $8\pi M^2 G_{\rm eff}|_{z=0}=G_N/G_{\rm eff,early} > 1 $, where we have set, as already stated before, $G_{\rm eff}|_{z=0}=G_N$, and used the limit $\lim_{z\to\infty}f_0'=1$ to define $G_{\rm eff,early}:=1/(8\pi M^2)$, which is equal to the gravitational constant felt on the cosmological background. This tendency can be directly seen by \eqref{fdS2}. In the limit $\bar{\rho}_{\rm m} \rightarrow 0$, namely in the far future, we obtain $\phi \rightarrow 0$ and then $f_0' \rightarrow 0$ which leads to $8\pi M^2 G_{\rm eff,future} \rightarrow \infty$. Because of this phenomenon, the model \eqref{fdS2} yields non-trivial dynamics which leads to a stop of the validity of the theory at some point in the evolution of the universe as $G_{\rm eff}$ becomes larger and larger. We also find that for this class of models we have $\lim_{\phi\to-\infty}w_{\rm DE}=-1/2$. Therefore, the dark energy component differs at early time from an effective cosmological constant.

More generally, from (\ref{eqn:eq_f'0}) one sees that $\bar{\rho}\to 0$ implies either $f'_0\to 0$ or $\phi\to 0$. If we demand that $G_{\rm eff}$ given by (\ref{eqn:Geff}) remains finite, then the only possibility is $\phi\to 0$, with which (\ref{eqn:eq_phi}) implies $H\to 0$. The same argument holds even without imposing $c_T=1$, based on (\ref{eq_phi})-(\ref{eqn:conservation}).

We therefore move on to the next model to describe a possibly interesting phenomenology for this theory.

\subsection{Dark energy with $w_{\rm DE}\ne -1$ from $\Lambda$}
We then consider the case $\Lambda \neq 0$ and assume
\begin{align}
f_0'=\frac{(M_*/M)^2+(\phi/\phi_c)^2}{1+(\phi/\phi_c)^2} \,, \label{case2}
\end{align}
with two constants $M_*$ and $\phi_c$. This model has the limits $f_0' \rightarrow 1$ for $|\phi| \gg |\phi_c|$ and $f_0' \rightarrow M_*^2/M^2$ for $|\phi| \ll |\phi_c|$. This model has two GR limits with different effective gravitational constants for $|\phi| \gg |\phi_c|$ and $|\phi| \ll |\phi_c|$. 

Eq.~\eqref{eq_phi} with $f_1=\sqrt{f_0'}$ gives the analytic solution for $\phi$,
\begin{align}
\frac{\phi^2}{\phi_c^2}=\frac{1}{2}\left( \frac{3 \bar{\rho}}{M^2 \phi_c^2} - \frac{M_*^2}{M^2}   \pm \sqrt{ \left( \frac{3 \bar{\rho}}{M^2 \phi_c^2} - \frac{M_*^2}{M^2}  \right)^2+  \frac{12 \bar{\rho}}{M^2 \phi_c^2} } \right)\,,
\end{align}
where we recall $\bar{\rho}=\bar{\rho}_{\rm m} + M^2 \Lambda$ is the total energy density including the cosmological constant. In the low energy limit $\bar{\rho} \ll M^2 \phi_c^2 $, two branches approach
\begin{align}
\phi^2 \rightarrow 0 ~(+\, {\rm branch}), \quad -M_*^2 \phi_c^2/ M^2 ~(-\, {\rm branch})\,,
\end{align}
and thus the minus branch is unphysical. We only consider the plus branch. The solution of $\phi$ shows that $\phi^2 \gg \phi_c^2$ for $\bar{\rho} \gg M^2 \phi_c^2 $ and $\phi^2 \ll \phi_c^2$ for $\bar{\rho} \ll M^2 \phi_c^2 $ which yield the early time gravitational constant $G_{\rm eff, early}=1/(8\pi M^2)$ and the late time one $G_{\rm eff, late}=1/(8\pi M_*^2)$, respectively. Depending on the value of $\phi_c$ this value for $G_{\rm eff}$ might be approximately reached even at $z\gtrapprox0$, and in this case we would have $G_{\rm eff,late}\approx G_N$. In any case, for high redshifts, we have $G_{\rm eff,z\to\infty}=1/(8\pi M^2)$. This means that the effective gravitational constant for the dark matter fluid fluctuations, at early times, reduces to the cosmological-background gravitational constant.

\begin{figure}[tbp]
\centering
\includegraphics[width=7cm,angle=0,clip]{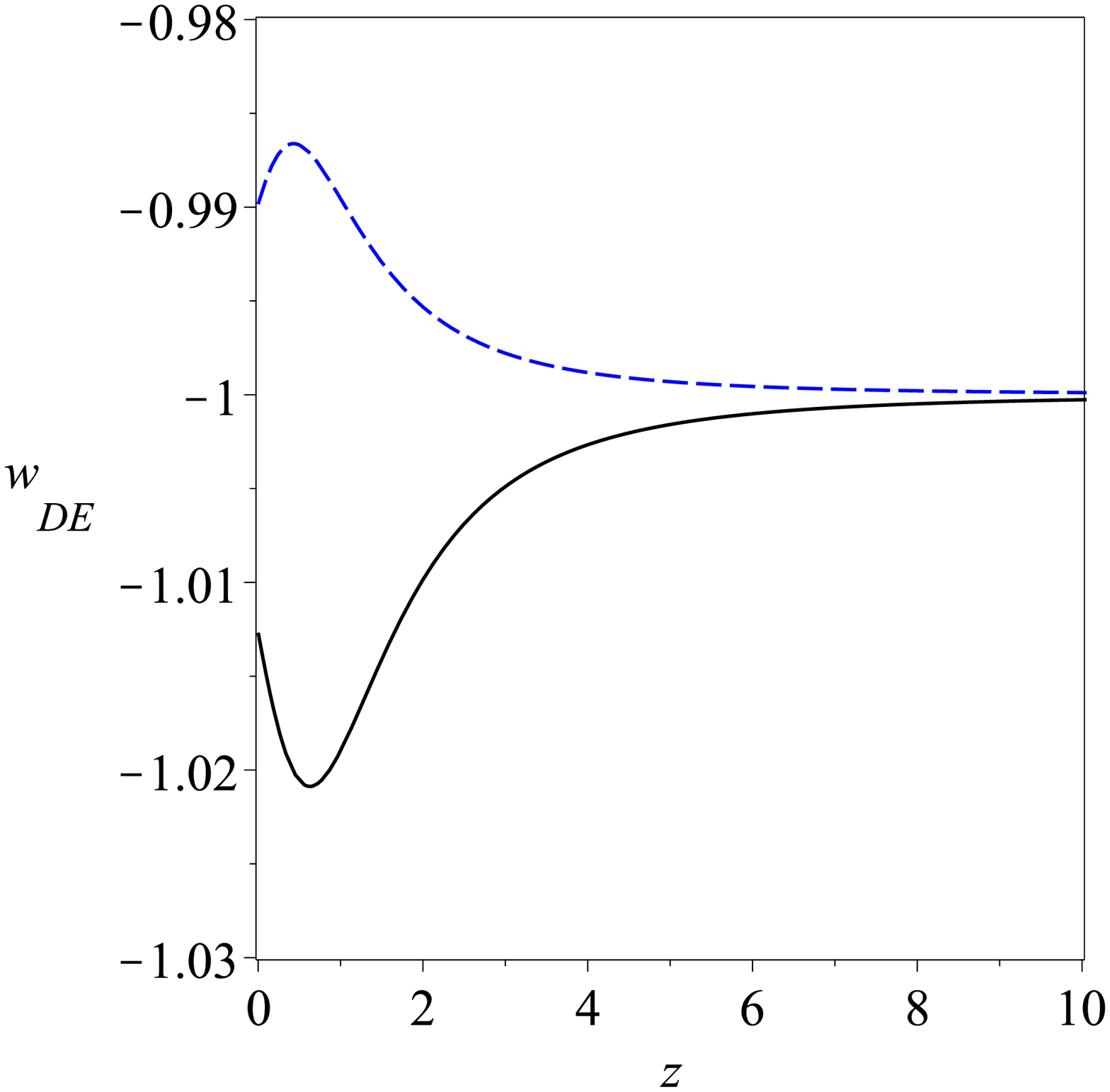}
\hspace{5mm}
\includegraphics[width=7cm,angle=0,clip]{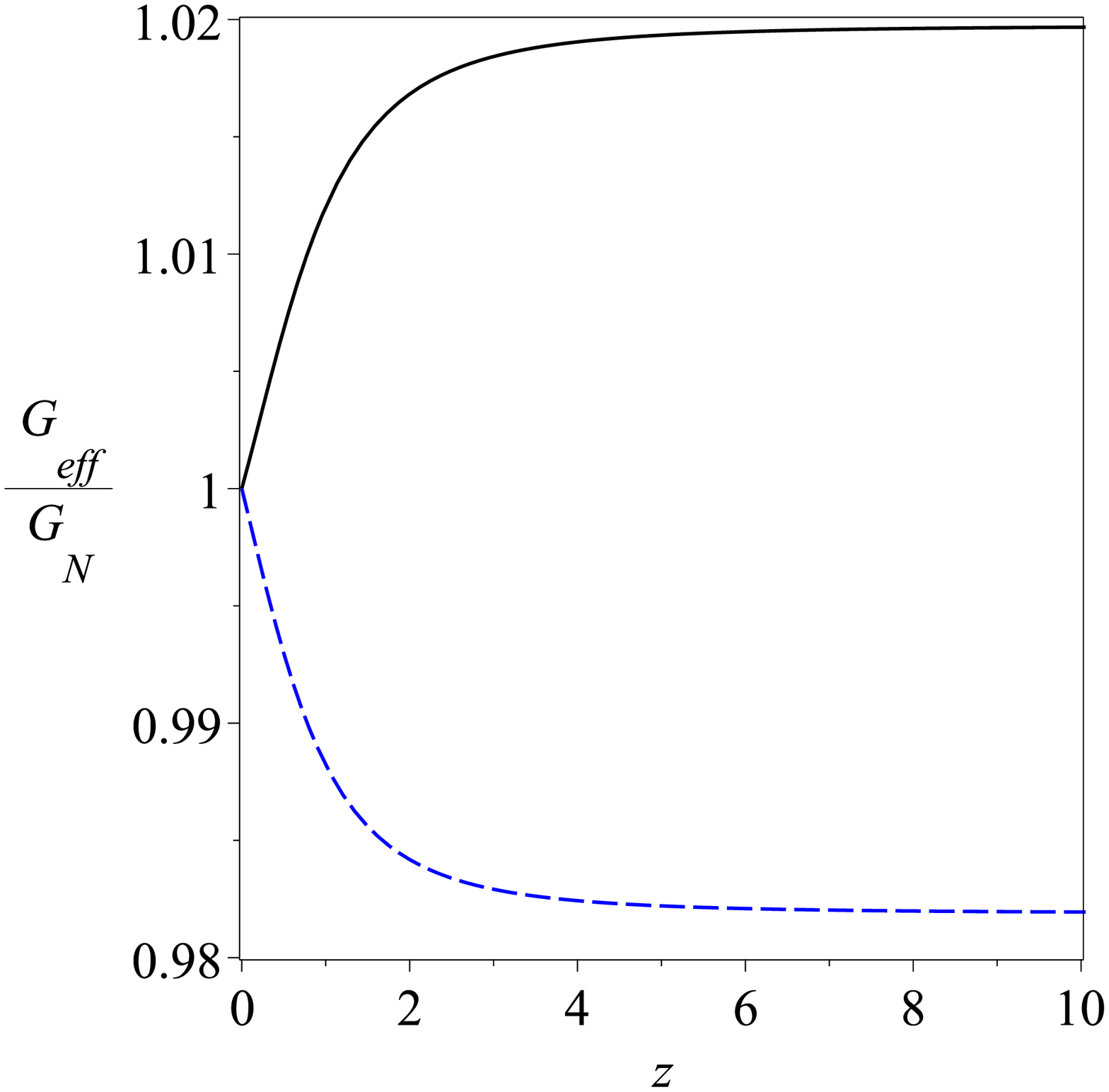}
\caption{Time evolutions of the equation of state parameter and the effective gravitational constant for the model \eqref{case2} with $M_*^2/M^2=1.1$ (black solid curves) and $M_*^2/M^2=0.9$ (blue dashed curves) and $\Lambda= \phi_c^2$. Here, we have also set $\Omega_{m0}=0.3$.
}
\label{fig_case2}
\end{figure}

However, on setting $\phi_c^2\simeq\Lambda\simeq H_0^2$, the modification from GR only appears in the present universe. Fig.~\ref{fig_case2} shows the time evolutions of $w_{\rm DE}$ and $G_{\rm eff}/G_N$ for the cases $M_*^2/M^2>1$ and $M_*^2/M^2<1$, respectively. Although the acceleration is caused by a cosmological constant, the effective equation of state parameter $w_{\rm DE}$ of dark energy is actually dynamical.

For this family of models one can also show that $\lim_{\phi\to-\infty}w_{\rm DE}=-1$. Therefore, with this model we are able to achieve early-times weak or strong gravity regimes for the dust-fluid fluctuations having a background cosmological constant at high redshifts. For this model to be viable we need to impose the background value of the gravitational constant, $1/(8\pi M^2)$ (which reduces to $G_{\rm eff,early}$, the effective gravitational constant felt by the dark matter perturbations at early times) not to be too different (up to a few percents) from $G_N$, the standard value for the Newton constant, which in this model corresponds to $G_{\rm eff}|_{z=0}$. If in the future data sets point more and more towards the possibility of early-times modifications of gravity, along the lines of \cite{Lin:2018nxe}, then this model will have a good chance to explain that behaviour.

\subsection{Reconstructed Dark Energy models}

Here we present the possibility of reconstructing the $f_0(\phi)$ function requiring the background to have a chosen dynamics. We will impose a dynamics for the background, for example
\begin{equation}
  H=H(a)=H_0\,E(a)\,, \label{eq:H}
\end{equation}
where $E$ is a given function of the scale factor which determines a particular dark energy dynamics, which satisfies the condition $E(a=1)=1$. In this case we also consider the presence of radiation, matter and a cosmological constant in terms of the matter content, so that we can introduce the Friedmann equation
\begin{equation}
  3M^2H_0^2E^2=\bar{\rho}_m+\bar{\rho}_r+\bar{\rho}_{\rm DE}\,,
  \end{equation}
where $\rho_r$ and $\rho_m$ are energy densities of radiation and matter, respectively, and $\bar{\rho}_{\rm DE}$ is given by (\ref{eqn:def-rhoDE}). From the equations of motion (\ref{eqn:eq_phi}) we can find
\begin{equation}
  f_0'' ={\frac { \left( 2\,\beta\,\bar\phi+6\,E \right) f_0'}{H_0\,\beta\,\bar\phi\, \left(3\,E -\beta\,\bar\phi \right) }}\,, \label{eq:recf02}
\end{equation}
where we have defined $\phi=\phi(a=1)\bar\phi$, and $\phi(a=1)=\beta H_0$. We find it useful to redefine the following constants
\begin{eqnarray}
 \Lambda&=&\lambda\,H_0^2\,,\\
  \bar{\rho}_{r0}&=&3M^2H_0^2\Omega_{r0}\,,\\
  \bar{\rho}_{m0}&=&3M^2H_0^2\Omega_{m0}\,,
\end{eqnarray}
so that $\bar{\rho}=M^2H_0^2(\lambda a^4+3\Omega_{m0}a+3\Omega_{r0})/a^4$ and that today we have
\begin{equation}
  \Omega_{m0}+\Omega_{r0}+\Omega_{{\rm DE}0}=1\,,
\end{equation}
where $\Omega_{\rm DE}:=\rho_{\rm DE}/(3M^2H^2)$. The equation of motion (\ref{eqn:eq_f'0}) also fixes $f_0'$ as
\begin{equation}
 f_0'= \frac{3(\lambda a^4+3\Omega_{m0}a+3\Omega_{r0})}{\bar{\phi}^2a^4\beta^2}\,.\label{eq:recf01}
\end{equation}
It is confirmed that the derivative of \eqref{eq:recf01} with respect to $\phi$ coincides with \eqref{eq:recf02} by using \eqref{eq_phi_cT1} and \eqref{eq:H}, and thus the reconstruction of $f_0'$ is self-consistent.
Finally, we can solve (\ref{eq_phi_cT1}) to find the dynamics of $\bar\phi$ at any time. By introducing the e-fold variable $\bar N:=\ln(a)$, the system of ODEs for this purpose is written as follows. 
\begin{eqnarray}
  \partial_N\bar\phi&=&{\frac {\bar\phi\, \left( 3\Omega_{m0}\,a+4\Omega_{r0} \right) 
               \left( \beta\bar\phi-3\,E \right) }{4E \left( \lambda\,{a}^{4}+3\,\Omega_{m0}\,a+3\Omega_{r0} \right) }}\,,\label{eq:ODEphi}\\
  \partial_N a&=&a\,,\\
        \bar\phi(\bar N=0)&=&1\,,\\
  a(\bar N=0)&=&1\,.
\end{eqnarray}
where we have introduced an ODE also for $a$ in order to keep the system autonomous. From Eq.\ (\ref{eq:recf01}), we find
\begin{equation}
  G_{\rm eff}=\frac1{8\pi M^2 f_0'}={\frac {{\bar\phi}^{2}{a}^{4}{\beta}^{2}}{24\pi M^2 \left( \lambda\,{a}^{4}+3\,\Omega_{m0}\,a+3\,\Omega_{r0} \right)}}\,,
\end{equation}
so that the Newton gravitational constant can now be written as
\begin{equation}
  G_{N}=G_{\rm eff}|_{\bar N=0}={\frac {{\beta}^{2}}{24\pi M^2 \left( \lambda+3\,\Omega_{m0}+3\,\Omega_{r0} \right)}}\,.
\end{equation}

After solving the system of ODEs, we are able to evaluate at any time the following observable
\begin{equation}
  \frac{G_{\rm eff}}{G_N}={\frac {(\lambda+3\,\Omega_{m0}+3\,\Omega_{r0})\,{\bar\phi}^{2}{a}^{4}}{ \lambda\,{a}^{4}+3\,\Omega_{m0}\,a+3\,\Omega_{r0} }}\,.
\end{equation}
For every selected dynamics $E(a)$ we will have a particular dynamics for the effective gravitational constant which can be tested, in principle, through several data sets. In general, constraints on $8\pi M^2 G_N$ will be coming from nucleosynthesis, whereas constraints for $G_{\rm eff}/G_N$ will come instead from the recombination time up to low redshift data, e.g.\ RSD experiments.

\section{Conclusion and Discussion}
\label{sec:summary}
Is GR the unique theory for the massless graviton with only two local degrees of freedom at low energy scale? Probably not. Even when the theory is completely equivalent to GR in the vacuum, the non-trivial and non-minimal coupling between matter field and gravity still yield non-standard predictions that differ from GR. In the present paper, by means of canonical transformations of lapse, 3-d induced metric as well as their conjugate momenta, we have generated such a type of gravity theory: one that is completely equivalent to GR in the absence of matter field, but in which gravity is nevertheless modified due to non-trivial matter coupling. We call this type of theory type-I minimally modified gravity theory, while type-II minimally modified gravity groups the ones in which there is no Einstein frame and therefore are inequivalent to GR even in the vacuum. The phenomenological deviation from GR is characterized by two functions $f_0(\phi)$ and $f_1(\phi)$ in the generating functional; for instance, the speed of GWs is given by $c_T^2=f_1^2/f_0'$. 

The Hamiltonian structure remains unchanged in the vacuum, however, the naive way of introducing a matter field breaks the temporal diffeomorphism and the Hamiltonian constraint gets downgraded to second class. One un-wanted degree of freedom appears in the phase space and renders the theory inconsistent. A consistent way to introduce matter coupling is to fix the gauge before coupling gravity to matter. In the present paper, we have employed the gauge fixing condition (\ref{eq:gauge-fix}). We have then introduced a k-essence field to represent a perfect fluid minimally coupled to the canonically transformed gravity theory. We find that in the short distance limit, the null energy condition for the k-essence matter field is required to avoid a ghost. The dispersion relation of scalar perturbations in the short distance limit is the same as the one in GR. 

It would be interesting to consider possible effects induced by quantum corrections. For example, since time-diffeomorphisms are no longer a symmetry of the theory, one may expect the appearance of new operators related to new massive states appearing at some high energy scale. As long as the theory is considered as a low energy effective field theory (EFT) below such a new cutoff, these new contributions can be taken as small. On another note, the observability of indirect Lorentz violations in the matter sector\textemdash for example those arising from graviton loops to matter propagators\textemdash could also be studied further. We however expect that such effects are also largely suppressed, since the Lorentz violation appears only at cosmological distances.

The deviation of the speed of GWs from that of photons is highly constrained by the recent observations of GWs at the relative error level of $10^{-15}$. We have thus considered the cosmology with $c_T^2=1$, i.e.\ $f_1^2=f_0'$. We have found that the canonical transformations of GR generates an effective dark energy term in the Friedmann equation even without including the bare cosmological constant in the first place. We have then performed a linear perturbation analysis and we find that the effective Newtonian constant equals the one in GR at present time, whereas it differs from it at early times. Therefore, the effective Newtonian constant smoothly reduces to the background gravitational constant, $1/(8\pi M^2)$, in the limit $f_0'\to1$. In fact, one will in general achieve either strong or weak gravity for the dark matter fluid perturbations at high redshifts. On the other hand, the slip parameter remains unchanged, i.e.\ $\eta\equiv \Psi/\Phi=1$. We have shown three models for which each one of them has a particular and distinguishable phenomenology, having a proper function of $f_0'$. In the case of non-zero bare cosmological constant, the equation of state is dynamical rather than constant $-1$ even though the cosmic acceleration is caused by a bare cosmological constant. Finally we were able to reconstruct the dynamics of $f_0'$ for a given dark energy dynamics $H=H(a)$.

In summary, we have proposed a systematic way to generate a type-I minimally modified gravity theory by a canonical transformation, following the general procedure of matter coupling introduced in \cite{Aoki:2018zcv}. The cosmic acceleration can be realized even without introducing a bare cosmological constant, while the speed of GWs and the effective Newtonian constant are compatible with current observations at low redshifts. Since the number of propagating degrees of freedom in the gravity sector is minimal, i.e.\ two, there is no need for a screening mechanism. In the present paper, we have focused on the late time cosmology. It would also be interesting to ask how the physics of the early universe is modified in this framework, provided that the EFT is still valid at high energy scale during the early universe. We may come back to this issue in the future.

\section*{Acknowledgments}

One of the authors (S.M.) acknowledges warm hospitality at the Henri Poincar\'{e} Institute and Institut Astrophysique de Paris, where a part of the work was completed. The work of K.A.~was supported in part by a Waseda University Grant for Special Research Projects (No.~2018S-128). A.D.F.\ was supported by JSPS KAKENHI Grant No.~16K05348. The work of C.L.\ is carried out under POLONEZ programme of Polish National Science Centre, No.~UMO-2016/23/P/ST2/04240,  which has received funding from the European Union's Horizon 2020 research and innovation programme under the Marie Sk\l odowska-Curie grant agreement No.~665778. The work of S.M.\ was supported in part by Japan Society for the Promotion of Science (JSPS)  Grants-in-Aid for Scientific Research (KAKENHI) No.~17H02890, No.~17H06359, and by World Premier International Research Center Initiative (WPI), MEXT, Japan. M.O.\ thanks Charles Dalang for his suggestions, and acknowledges the support from the Japanese Government (MEXT) Scholarship for Research Students. 
 \includegraphics[width=0.08\textwidth]{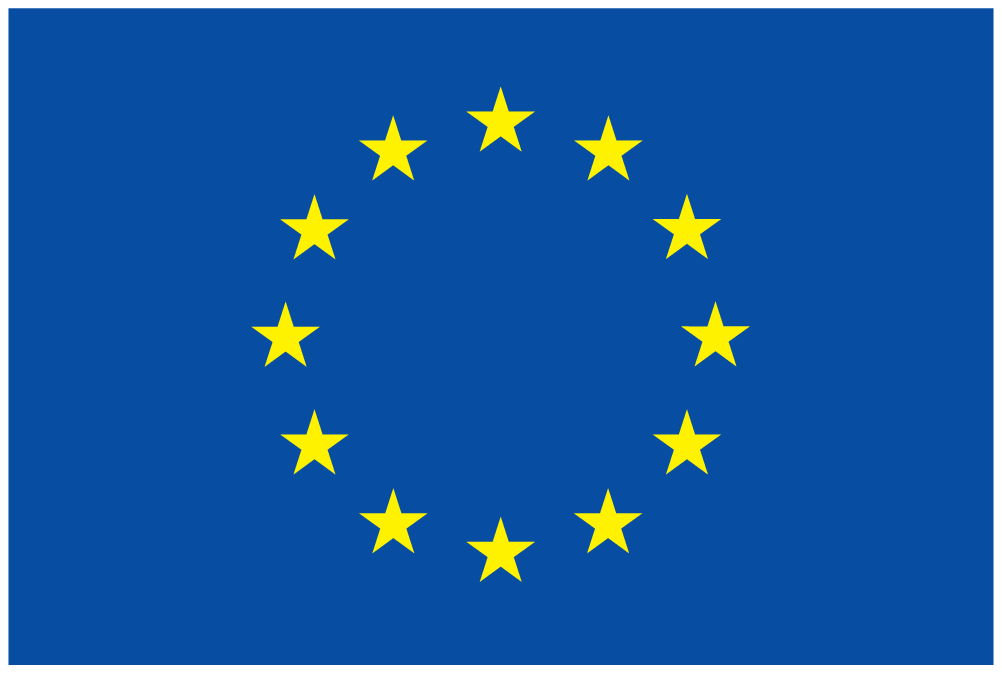}


\bibliography{ref181206}
\bibliographystyle{JHEP}

\end{document}